\documentclass[a4paper,20pt]{article}
    \usepackage[T1]{fontenc}
    \usepackage{graphicx}
    \usepackage{epsfig}
    \usepackage{amsmath}
    \usepackage{amsfonts}
    \renewcommand{\abstract}{}
\begin{document}
\makeatletter
\renewcommand{\@oddhead}{\textit{YSC'14 Proceedings of Contributed Papers} \hfil \textit{T. Kisiel, M. Soida}}
\renewcommand{\@evenfoot}{\hfil \thepage \hfil}
\renewcommand{\@oddfoot}{\hfil \thepage \hfil}
\fontsize{11}{11} \selectfont

\title{MHD Space Sailing}
\author{\textsl{T. Kisiel$^{1}$, M. Soida$^{2}$}}
\date{}
\maketitle
\begin{center} {\small $^{1}$Jan D\l ugosz Academy, Institute of Physics, al. Armii
Krajowej 13/15, 42-200 Cz\k{e}stochowa, Poland \\
$^{2}$Astronomical Observatory of the Jagiellonian University, ul. Orla 171, 30-244 Krak\'ow, Poland \\ t.kisiel@ajd.czest.pl, soida@oa.uj.edu.pl}
\end{center}

\begin{abstract}
The rocket technology dates back as far as medieval China.
Used initially for entertainment and religious practices over time rockets
evolved into weapons and finally into means of transportation.

Today, we are nearing the top of the rockets' capabilities. Although,
for now they are the only way for us to send anything into space
we are becoming more and more aware of the limitations of this technology.
It is essential that we invent other means of
propelling probes and other interplanetary vehicles through space.

The authors had performed a series of magnetohydrodynamic
simulations using the University of Chicago's Flash package to find
out whether the interactions between the Solar Wind and the conducting
ring with the electric current would occur. The MHD simulations gave
the results similar to the monte-carlo calculations performed by dr
Charles Danforth [1] from the University of Colorado. It is the
authors' conclusion that the promising results should encourage
further study of the phenomenon and the possibility of using it in practice.
\end{abstract}

\section*{Introduction}

\indent \indent Rockets were known already in Medieval China.
They were initially used during the religious rituals. Later they
became primitive weapons. Since that time we developed technologies
which allowed us to turn rockets into the only mean of sending
manmade objects such as probes into far reaches of the Solar System.

The principle is simple - gain enough momentum in the first
few minutes to break free from the Earth's gravity and inertially
reach the desired target. The process requires lots of energy and
consumes great ammounts of fuel. Two limitations become apparent.
Firstly, we are limited by the mass of the object we want to send.
The heavier it is, the more fuel we need; the more fuel we have,
the heavier the rocket and the more fuel it needs to lift itself.
It seems that we reached mass limits in the 1960's and 70's when
the biggest rockets such as the Saturn V were built.
Secondly, once we start our journey, there's no turning back.

It is necessary, then, to find alternatives to rockets.
Once we move outside the Earth's magnetosphere we could, for example,
use the Solar Wind for propulsion.

In this paper we are presenting initial results of the MHD
simulations performed in the Astronomical Observatory in Krak\'ow,
testing the interactions of the uniform flow of the plasma stream
(representing the Solar Wind)
and the magnetic field produced by a conductive ring with
the electric current flowing through it. The simulations were
made with the use of the FLASH 2.3 package written at the
University of Chicago. The results are very interesting and
hopefuly will lead to further studies of the phenomenon.

\section*{The model}

\indent \indent In this paper we're using the simplest dipole
magnetic field produced by a single loop of the conductive material.

The analytical methods can describe the electromagnetic potential $A$ around
the loop with the following equation [2]:

\begin{equation}
A_{\phi}(r, \theta )= \frac{4Ia}{c\sqrt{a^2+r^2+2ar\sin\theta}}\left[ \frac{(2-k^2)K(k)-2E(k)}{k^2}\right],
\end{equation}
with E and K being the elliptical integrals of $k$:
\begin{equation}
k^2=\frac{4ar\sin\theta}{a^2+r^2+2ar\sin\theta}.
\end{equation}

One can only solve those analytically for small $k^2$ which is only valid
when $a\gg r$, $a\ll r$ or $\theta\ll 1$. In these simulations
we need a solution which would be valid
for the whole area. We decided to use an approximated electromagnetic
potential which properties would be similar to the numerical solution
but which would be described with a single equation.

\indent After some research we decided to use the following equation
which we found to be a good estimate of the numerical solutions of
the potential above:

\begin{equation}
A_{\phi}(\rho , z, \phi)=C\frac{R\rho}{20R^2+z^2}\exp{\left(1-\frac{2z^2}{R^2}-\frac{20R\rho}{20R^2+z^2}\right)},
\end{equation}
where $R$ is the loop's radius and $C$ is a constant containing
all the physical constants. To simplify the equation the formula was
chosen in the cylindrical coordinates. This form of the electromagnetic potential works best for our simulations.

\section*{FLASH}
\indent \indent The FLASH is a simulation package created by the University of Chicago. It
is written mainly in Fortran 90 and the structure allows for easy modifications of the code
to suit the needs of the user. It can be used for a wide range of numerical simulations
although its main function is to simulate hydrodynamical and magnetohydrodynamical problems~[3].

As the initial conditions we used the magnetic field equations derived directly
from the above potential equation. The initial simulations resulted in the magnetic field
being carried away by the plasma outside of the simulated area. We then modified the code
to freeze the evolution of the magnetic field. Unfortunately, by doing it this way we lost
information about the energy required to maintain the magnetic field but our goal was only
to simulate forces experienced by the loop resulting from the pressure and density changes
of the plasma stream leaving the energy considerations for future research.

\section*{Results}

\indent \indent We transformed the initial results to correspond to the folowing parameter
values:
\begin{itemize}
\item the electic current $I=1kA$
\item the radius of the loop $R=2m$
\item the length of the simulation are cube $l=120m$
\item the plasma stream velocity $v=400km/s$ \,[4]
\item the plasma stream density $rho=10 particles/cm^3$ \,[4]
\end{itemize}

We made the simulations for a series of pitch angles. Of course, in space we can only
change the position of the loop but it was much easier to change the  direction of
the stream in the code. In each of the simulated cubes the loop was in the XY plane while
the stream was coming at the angle to the XY plane.

The first and obvious result is that there is an interaction between the plasma stream and
the magnetic field. That was to be expected. We then tried to calculate the overall
momentum change of the stream which would tell us if we can expect any `thrust' from the
loop or not. We calculated the resultant momentum vector in the IDL package.
Table 1 shows the value and direction of the resulting force.

The interesting result is that there is a non-parallel force component. At first it may seem
to be a violation of a momentum conservation principle but it isn't. It is a result of an
asymetry of the magnetic field when the stream flows with pitch angles different then 0 or
90 degrees. The asymetry makes different parts of the stream to have different lenghts of
their trajectories around the loop. This is very similar to the lift force resulting from
the air flowing around a wing of a plane.

\begin{table}[!ht]
\begin{minipage}[b]{.95\linewidth}
\centering \caption{The components of the force vector per volume
unit in $10^{-3} N/m^3$.}\vskip 5mm \centering
\begin{tabular}{|c|c|c|c|c|}
\hline
angle & $F_x/V$ & $F_y/V$ & $F_z/V$ & |F|/V\\
\hline
0&-3.24 & -8.562$\cdot10^{-6}$ & 1.141$\cdot$ $10^{-6}$ & 3.24\\
30&-10.83 & 4.866$\cdot10^{-6}$ & -2.24 & 11.06\\
60 & 19.28 & 1.226$\cdot10^{-4}$ & -31.51 & 36.94\\
90 & 0.212 & -1.678$\cdot10^{-5}$ & -50.69 & 50.69\\
\hline
\end{tabular}
\end{minipage}
\end{table}

\section*{Conclusions}

\indent \indent The first conclusion is that there is indeed a force which could be used to
move objects through space. Although, the force is very small (to compare, a single engine
of a Boeing 737 jet plane gives as much as about 100 kN of thrust) we studied the simplest
configuration of magnetic field and its source. With stronger fields and larger areas it
would cover the force may increase dramatically.

Moreover, we completely neglected the reconnection of the magnetic
field. It is known that the magnetic field is carried by the Solar
Wind. The pressure of the magnetic field itself may be significant
but that requires more studies.

The main conclusion here should be that the entire phenomenon
relatively unresearched. The initial results of the few teams which
has studied it are promising. They should encourage other
researchers to study the problem in much more details than presented
here. It also promises that the phenomenon could be used
successfully in practice.

\section*{Acknowledgements}
\indent \indent We would like to thank dr Bogdan Wszo\l ek for
making it possible to present this paper on the YSC conference in
Kiev. Also, we would like to thank everyone who shared their
thoughts and ideas which were invaluable during the initial stages
of the simulations. 
\newpage
\textbf{Figure 1.} A simple loop of a radius $a$ with the electric
current $I$ flowing through it.\vspace{10ex}

\textbf{Figure 2.} Vectors of magnetic field $\vec B$ in the XZ
plane.\vspace{10ex}

\textbf{Figure 3.} Density of the plasma stream with the relative
pitch angle between the stream and the loop of 0, 30, 60 and 90
degrees.\vspace{10ex}

Figures are available on YSC home page
(http://ysc.kiev.ua/abs/proc14$\_$9.pdf).

\end{document}